\documentclass{lhep}       

\journal{LHEP}
\vol{xx}
\jyear{2020}
\pages{xxx} 

\usepackage[utf8]{inputenc}
\usepackage{graphics,float}
\usepackage{graphicx}
\usepackage{color}
\usepackage{amssymb,amsfonts,amsmath}
\usepackage{xspace}
\usepackage{placeins}
\usepackage{dcolumn}
\usepackage{bm}
\usepackage{longtable}
\usepackage{pbox}

\usepackage{hyperref}
\usepackage{indentfirst}  
\usepackage{rotating} 
\usepackage{longtable,multirow}

\newcommand{\met}{\ensuremath{E_{T}^{\text{miss}}}}

\def\lsim{\mathrel{\raise.3ex\hbox{$<$\kern-.75em\lower1ex\hbox{$\sim$}}}}
\def\gsim{\mathrel{\raise.3ex\hbox{$>$\kern-.75em\lower1ex\hbox{$\sim$}}}}
\def\ifmath#1{\relax\ifmmode #1\else $#1$\fi}

\def\Ztwo{\ensuremath{\mathbb{Z}_2}}

\def\BR{{\rm{BR}}}

\def\eff{\epsilon}
\def\code#1{{\tt{#1}}}

\begin{document}

\title{SModelS database update v1.2.3}

\author{Charanjit K. Khosa,\auno{1} 
Sabine~Kraml,\auno{2} 
Andre~Lessa,\auno{3} 
Philipp Neuhuber,\auno{4} and 
Wolfgang~Waltenberger\auno{4}}
\address{$^1$ Department of Physics and Astronomy, University of Sussex, Brighton BN1 9QH, UK}
\address{$^2$Laboratoire de Physique Subatomique et de Cosmologie, Universit\'e Grenoble-Alpes, CNRS/IN2P3, Grenoble INP,\\ 53 Avenue des Martyrs, 38000 Grenoble, France}
\address{$^3$Centro de Ci\^encias Naturais e Humanas, Universidade Federal do ABC, Santo Andr\'e, 09210-580 SP, Brazil}
\address{$^4$Institut f\"ur Hochenergiephysik,  \"Osterreichische Akademie der Wissenschaften, Nikolsdorfer Gasse 18, 1050 Wien, Austria}

\begin{abstract}
We present an update of the SModelS database with simplified model results from 13~ATLAS and 10~CMS searches 
for supersymmetry at Run~2. This includes 5~ATLAS and 1~CMS analyses 
for full Run~2 luminosity, i.e.\ close to 140~fb$^{-1}$ of data. In total, 76 official upper limit and efficiency map results have been 
added. Moreover, 21 efficiency map results have been produced by us using MadAnalysis5,  
to improve the coverage of gluino-squark production. 
The constraining power of the new database, v1.2.3, is compared to that of the previous release, v1.2.2.  
SModelS~v1.2.3 is publicly available and can readily be employed for physics studies.
\end{abstract}

\maketitle

\begin{keyword}
physics beyond the standard model\sep supersymmetry\sep simplified models\sep  LHC\sep reinterpretation
\end{keyword}

\section{Introduction}\label{intro}

\href{https://smodels.github.io/}{SModelS}~\cite{Kraml:2013mwa,Ambrogi:2017neo,Ambrogi:2018ujg} 
is a public software tool that enables the fast interpretation of simplified model results from ATLAS and CMS searches for 
supersymmetry (SUSY) in an automatized way. 
It can be used for evaluating the collider signals of any Beyond the Standard Model (BSM) scenario with a \Ztwo-like symmetry, 
for which the signal acceptance of the SUSY searches apply~\cite{Kraml:2013mwa}. 
The working principle of SModelS is to decompose all signatures occurring in a given model or scenario into simplified model  topologies---also referred to as simplified model spectra (SMS)---by means of a generic procedure where each topology is defined by the vertex structure, the Standard Model (SM), and BSM final states;
intermediate \Ztwo-odd BSM particles are characterized only by  their  masses,  production  cross  sections,  and  branching ratios.

The signal weights, determined in terms of cross sections times branching ratios, $\sigma\times\BR$, are then matched against a database of LHC results. This is easier and much faster than reproducing analyses with Monte Carlo event simulation, and it allows for reinterpreting searches which are not just cut and count, e.g., analyses which rely on BDT (boosted decision tree) variables.
The downside is that the applicability is limited by the simplified model results available in the database.
Moreover, whenever the tested signal splits up into many different channels, as often the case in complex models with many new particles, the derived limits tend to be highly conservative.  
SModelS is thus particularly useful for evaluating constraints and generally characterizing collider signatures in 
large scans and model surveys. 

SModelS makes use of two types of experimental results: upper limit (UL) results and efficiency map (EM) results. 
Upper limit results provide 95\% confidence level (CL) upper limits on $\sigma\times\BR$ as a function of the respective 
parameter space of the simplified model---usually BSM masses or slices over mass planes. Their advantage is that  
the statistical evaluation (i.e., combination of signal regions  when relevant, limit setting procedure, etc.) is done directly by the experimental collaboration. Furthermore, limits obtained from non-cut and count analyses can also be used.
However, their statistical interpretation is limited, only allowing for an excluded or not statement, on a purely topology-per-topology basis. 
Only if the {\it expected} UL maps are also available, it becomes possible to select the most sensitive result and/or to compute an approximate likelihood for the signal strength as a truncated Gaussian~\cite{Azatov:2012bz}. 
Efficiency maps correspond to grids of simulated acceptance times efficiency  ($A\times\eff$) values (simply called `efficiencies' in the following) for the various signal regions of an analysis.  
Their advantage is that they allow for combining contributions from different simplified model topologies to the same signal region, and for computing the likelihood~\cite{Ambrogi:2017neo,Ambrogi:2018ujg}. 

Besides speed, the power of SModelS comes from its large database of results, which is regularly updated. 
For Run~1, not counting superseded results, the SModelS database contains 93 official UL and 72 EM results from 17 ATLAS and 18 CMS analyses.  
In \cite{Dutta:2018ioj}, we presented the implementation of the Run~2 SUSY search results from CMS with 36~fb$^{-1}$ from the Moriond and the summer (LHCP and EPS) conferences of 2017; this amounted to 84 new UL maps from 19 different analyses.  
This was further augmented by CMS long-lived particle (HSCP and R-hadron) constraints \cite{Heisig:2018kfq} and the first set of 12 UL maps from six ATLAS SUSY analyses with 36 fb$^{-1}$ in  \cite{Ambrogi:2018ujg}.  
Moreover, we included 30 ``home-grown'' EM results relevant for constraining gluino-squark production at 8~TeV~\cite{Ambrogi:2018ujg}.  

In the new v1.2.3 presented in this paper, this is extended by 
the simplified model results from 13 new ATLAS and 10 CMS searches for SUSY at Run~2, including a first set of results for full Run~2 luminosity. 
In total, 76 official UL and EM results have been added. 
Moreover, 21 EM results (consisting of 351 individual EMs when counting each maps in each signal region separately) for 13~TeV were produced by us for better covering scenarios where gluino-squark associated production is important. 
In the following, we discuss in detail which results have been added and compare the 
constraining power of the new database, v1.2.3, to that of the previous release, v1.2.2.

\begin{table*}[t!]
\tbl{Official ATLAS 13~TeV results included in this SModelS database update. All analyses assume large \met\ in the final state. 
The 5th column lists the specific SMS results included, using
the shorthand ``txname'' notation (see text for details).
For brevity, only the on-shell results are listed, although the off-shell ones are always also included
(e.g., T2tt in the table effectively means T2tt and T2ttoff; see also \cite{Dutta:2018ioj}). 
The superscript $\ddag$ denotes the topologies, for which official EMs from ATLAS are included in addition to the UL maps.  
The superscript $*$ denotes UL results which contain expected limits in addition to the observed ones. 
\label{tab:ATLAS}}{
\begin{tabular}{ lcclll }
{\bf Analysis}  &  {\bf\bm $\cal{L}$ [fb$^{-1}$]}  & {\bf Ref.} & {\bf ~~~~ID} & {\bf SMS results (txnames)} & {\bf Type} \\ \hline
0 lept.\ + jets   & 36.1 & \cite{Aaboud:2017vwy} & SUSY-2016-07 & T1, T2, T5WW, T5ZZ, T5WZh, T6WW, T6WZh & UL \\
0 lept.\ stop & 36.1 & \cite{Aaboud:2017ayj} &  SUSY-2016-15 & T2tt, T2bbffff & UL \\
1 lept.\ stop & 36.1 & \cite{Aaboud:2017aeu} &  SUSY-2016-16 & T2tt$^\ddag$, T2bbffff$^\ddag$, T6bbWW$^\ddag$ & UL, EM \\
2--3 lept.\  & 36.1 & \cite{Aaboud:2018jiw} &  SUSY-2016-24 & TChiWZ$^\ddag$,  TSlepSlep$^\ddag$, & UL, EM \\
				& & & & TChiChipmSlepSlep, TChipChimSlepSlep & UL \\
photon + jets & 36.1 & \cite{Aaboud:2018doq} &  SUSY-2016-27 & T5gg$^\ddag$, T5Zg, T6gg$^\ddag$, TChipChimgg & UL, EM \\
0--1 lept. + $b$-jets  & 36.1 & \cite{Aaboud:2017wqg} &  SUSY-2016-28 & T2bb & UL \\
 EW-ino, Higgs & 36.1 & \cite{Aaboud:2018ngk} &  SUSY-2017-01 & TChiWH & UL \\
 Higgsino, Z/H & 36.1 & \cite{Aaboud:2018htj} &  SUSY-2017-02 & TChiH & UL \\
 2 OS taus & 139.0 & \cite{Aad:2019byo} &  SUSY-2018-04 & TStauStau & UL \\
 3 lept., EW-inos & 139.0 & \cite{Aad:2019vvi} &  SUSY-2018-06 & TChiWZ & UL$^*$ \\
multi-$b$ & 139.0 & \cite{Aad:2019pfy} &  SUSY-2018-31 & T6bbHH & UL \\
 2 OS lept. & 139.0 & \cite{Aad:2019vnb} &  SUSY-2018-32 & TChiWW, TChipChimSlepSlep, TSlepSlep & UL \\
 1 lept. + $H\to b\bar b$ & 139.0 & \cite{Aad:2019vvf} &  SUSY-2019-08 & TChiWH & UL \\
\hline
\end{tabular}}
\end{table*}

\begin{table*}[t!]
\tbl{Official CMS 13~TeV results included in this SModelS database update; see caption of Table~\ref{tab:ATLAS} for details.  
\label{tab:CMS}}{
\begin{tabular}{ lcclll }
{\bf Analysis}  &  {\bf\bm $\cal{L}$ [fb$^{-1}$]}  & {\bf Ref.} & {\bf ~~~~ID} & {\bf SMS results (txnames)} & {\bf Type} \\ \hline
0 lept., top tagging   & 2.3 & \cite{Khachatryan:2017rhw} &  SUS-16-009 & T1tttt, T2tt, T5tctc & UL$^*$ \\
$\geq 2$ taus & 35.9 & \cite{Sirunyan:2018vig} &  SUS-17-003 & TChiChipmStauStau, TChipChimStauSnu & UL \\
EW-ino combination  & 35.9 & \cite{Sirunyan:2018ubx} &  SUS-17-004 & TChiWH, TChiWZ & UL \\
1 lept. compressed stop  & 35.9 & \cite{Sirunyan:2018omt} &  SUS-17-005 &  T2bbffff, T6bbWWoff & UL$^*$ \\   
				jets + boosted $H\to b\bar b$  & 35.9 & \cite{Sirunyan:2017bsh} &  SUS-17-006 & T5HH, T5HZ & UL$^*$ \\
2 SFOS lept. & 35.9 & \cite{Sirunyan:2018nwe} &  SUS-17-009 & TSelSel, TSmuSmu, TSlepSlep & UL$^*$ \\
2 lept. stop & 35.9 & \cite{Sirunyan:2018lul} &  SUS-17-010 & T2tt, T6bbWW, TChipChimSlepSnu & UL$^*$ \\
photon + ($b$) jets  & 35.9 & \cite{Sirunyan:2019hzr} &  SUS-18-002 & T5Hg, T5bbbbZg, T5ttttZg, T6ttZg & UL$^*$ \\
0 lept. + jets, MHT    & 137.0 & \cite{Sirunyan:2019ctn} &  SUS-19-006 & T1, T1bbbb, T1tttt, T2, T2bb, T2tt & UL$^*$ \\
\hline
\end{tabular}}
\end{table*}

\begin{table*}[t!]
\tbl{Home-grown 13~TeV EM results included in this SModelS database update. Again, for brevity only on-shell results are listed, but  the off-shell ones are included as well (this concerns T1tttt, T2tt, T5WW and T6WW; see \cite{Dutta:2018ioj} for details). 
\label{tab:homegrown}}{
\begin{tabular}{ lcclll }
{\bf Analysis}  &  {\bf\bm $\cal{L}$ [fb$^{-1}$]}  & {\bf Ref.} & {\bf ~~~~ID} & {\bf SMS results (txnames)} & {\bf Type} \\ \hline
0 lept.\ + jets   & 36.1 & \cite{Aaboud:2017vwy} & ATLAS-SUSY-2016-07 & T1, T2, TGQ, T3GQ, T5GQ, T5WW, T5ZZ, T6WW &  EM \\
0 lept. + jets  & 35.9 & \cite{Sirunyan:2017cwe} &  CMS-SUS-16-033 & T1, T1bbbb, T1tttt, T2, T2bb, T2tt, TGQ, T3GQ, T5GQ & EM \\
\hline
\end{tabular}}
\end{table*}

The codebase of SModelS~v1.2.3 is almost identical with v1.2.2: apart from two minor bug fixes, the most relevant change is that the database is now by default
downloaded from \code{smodels.github.io}, contrary to \code{smodels.hephy.at} in v1.2.2; see the release notes at  \url{https://smodels.readthedocs.io/en/stable/ReleaseUpdate.html}.

Throughout this paper, we assume that the reader is already familiar with SModelS. If this is not the case, the reader may find detailed explanations about the usage and inner workings of SModelS in~\cite{Kraml:2013mwa,Ambrogi:2017neo,Ambrogi:2018ujg} and the online manual at \url{https://smodels.readthedocs.io}.
A detailed discussion of related approaches is given in \cite{Abdallah:2020pec}.

\section{New results in the database}


The new (official) ATLAS and CMS results included in the v1.2.3 database are detailed in Tables~\ref{tab:ATLAS} and \ref{tab:CMS}. They concern all applicable new SUSY search results 
since SModelS v1.2.2, for which simplified model results are available in digital form on HEPData or the analysis' TWiki page (status end of March 2020). Note that six analyses, five from ATLAS and one from CMS, are for full Run~2 luminosity of about 140~fb$^{-1}$. The home-grown EM results produced by us are listed in Table~\ref{tab:homegrown}; we will come back to these later. 

Inside SModelS, individual SMS results are identified by the analysis ID and the ``txname'', which describes in a shorthand notation the hypothesized SUSY process (largely following \cite{Chatrchyan:2013sza}). 
Due to lack of space we do not elaborate this naming scheme here,  but refer the reader to our ``SMS Dictionary'' at \url{https://smodels.github.io/docs/SmsDictionary123}, which provides a complete list of txnames together with the corresponding diagrams.  
Each included map is thoroughly validated to make sure that it reproduces the limits reported in the experimental publication. 
Detailed validation plots for each result are available online at  \url{https://smodels.github.io/docs/Validation123}.  

\begin{figure*}[t!]\centering
\includegraphics[height=2.5cm]{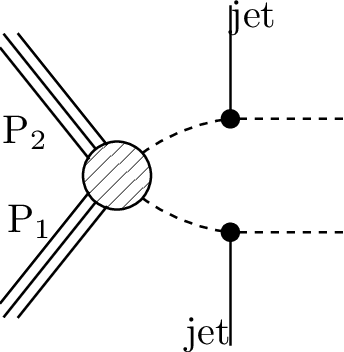}\qquad 
\includegraphics[height=2.5cm]{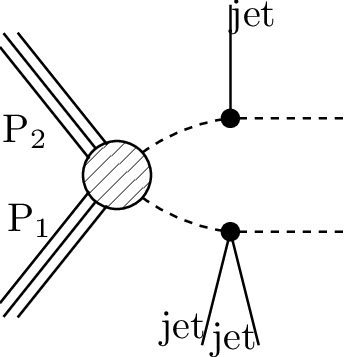}\qquad
\includegraphics[height=2.5cm]{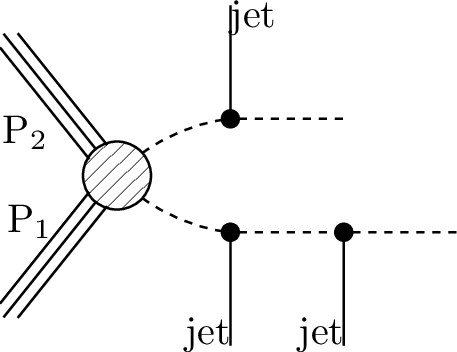}\qquad
\includegraphics[height=2.5cm]{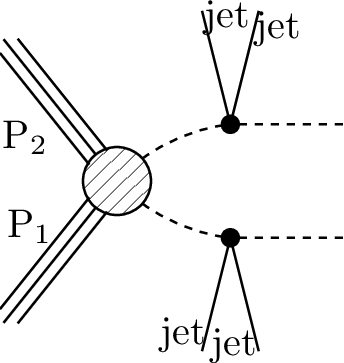}\qquad
\includegraphics[height=2.5cm]{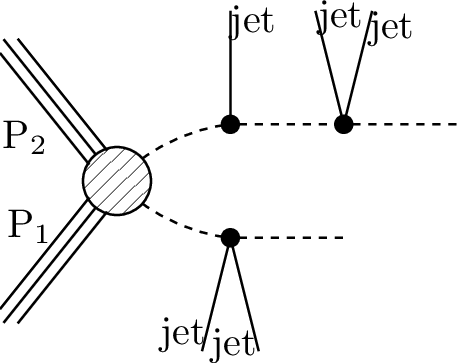} 
\caption{\label{fig:EMbakery}
Sample of SMS topologies with 2--5 jets relevant for gluino/squark production, for which ``home-made'' EM results have been produced with the MadAnalysis\,5 recast codes \cite{ma5recast:altas,ma5recast:cms} for the ATLAS and CMS 13~TEV multi-jet + \met\ searches \cite{Aaboud:2017vwy,Sirunyan:2017cwe}. 
From left to right: T2, TGQ, T3GQ, T1, and T5GQ.}
\end{figure*}

A couple of comments are in order regarding the SMS (UL and EM)  
results in Tables~\ref{tab:ATLAS} and \ref{tab:CMS}. 
First, with the exception of ATLAS-SUSY-2018-06~\cite{Aad:2019vvi}, the UL maps provided by ATLAS contain only the observed limits but not the expected ones. This makes it impossible to deduce the statistically most sensitive analysis, or to estimate a CL for a given hypothesized signal. 
On the CMS side, SUS-17-003 and SUS-17-004 do not report expected limits.   

Second, while we \emph{very highly} appreciate the provision of EMs by ATLAS, in many cases, the $A\times\eff$ values are for the most sensitive (a.k.a.\ ``best'') signal region only. This is not optimal because the best signal region (at a given mass point) can depend on the tested signal and hence vary for different BSM scenarios. We, therefore, want to encourage ATLAS and CMS to provide EMs for all signal regions over the full parameter space of the considered simplified model. In case the number of signal regions is too large, as often the case for CMS searches, this might be done for a set of appropriately aggregated regions.    

Third, some EMs provided by ATLAS were not included in the v1.2.3 database because they apply to a sum over SMS topologies instead of a single topology. This is, in particular, the case when mixed decays are assumed, e.g., $\tilde\chi^0_2\to \tilde\chi^0_1 + Z$ or $h^0$ with fixed BRs, or $\tilde\chi^\pm$ decays via sleptons, with the contributions of charged sleptons and sneutrinos summed over. 
From these combined efficiencies, there is not enough information to infer the contribution from each individual topology.
Hence, the result cannot be applied to general models, where the topologies can contribute with different relative weights. 
Such results are, therefore, not included in SModelS.

A general discussion of simplified model results and  recommendations regarding their presentation 
can be found in  Section~II.A.5 of the recent report of the LHC Reinterpretation Forum~\cite{Abdallah:2020pec}. 

Concerning the second point above, we note that ATLAS has started to release full likelihoods in the form of a \code{json} serialization~\cite{ATL-PHYS-PUB-2019-029}, which should describe background correlations at the same fidelity as the likelihood model used in the experiment. So far this is available for two analyses, the sbottom multibottom search~\cite{Aad:2019pfy} (SUSY-2018-31) and the search for direct stau production~\cite{Aad:2019byo} (SUSY-2018-04), both for full Run~2 luminosity. Both analyses also provide detailed EMs for the simplified models they consider. 
This is a pioneering step forward in the presentation of BSM searches, which can greatly improve the preservation and reuse of the experimental results. 
We are currently working on an interface to \code{pyhf}~\cite{pyhf} to make full use of these data (see contribution no.~15 in \cite{Brooijmans:2020yij}). 
In the meanwhile, the TStauStau and T6bbHH EMs from 
\cite{Aad:2019byo,Aad:2019pfy} are not included in the new SModelS v1.2.3 database, because using only the best signal regions for limit setting leads to over- or underexclusions compared to the UL results --- we think that this is due to background fluctuations and, thus, postpone the usage of the TStauStau and T6bbHH EMs until the full \code{json/pyhf} implementation is available in SModelS. 


Let us now turn to the home-grown EMs in Table~\ref{tab:homegrown}.
For a good coverage of complex models with several new particles, it is crucial that all major contributions to the total signal  considered by a particular analysis can be taken into account. 
In the SUSY context, this means that a large set of EM results is required in particular for the generic gluino/squark searches;  see~\cite{Ambrogi:2017lov,Chalons:2018gez}. Concretely, it is important to cover topologies arising from gluino-squark associated production in addition to the usual gluino-pair and squark-pair productions~\cite{Ambrogi:2017lov}. 

We, therefore, turned to recast the ATLAS and CMS multijet + \met searches \cite{Aaboud:2017vwy,Sirunyan:2017cwe} with 
MadAnalysis\,5~\cite{Conte:2012fm,Dumont:2014tja,Conte:2018vmg} in order to produce such a set of EMs for 13~TeV which 
were then incorporated into the SModelS v1.2.3 database. 
(Similar home-grown EMs for 8~TeV were already included in v1.2.2~\cite{Ambrogi:2018ujg}.) We considered topologies with 2--5 jets in the final state as shown in Figure~\ref{fig:EMbakery}, 
to allow for the combination of gluino-pair, squark-pair, and gluino-squark associated production. 
This was augmented with EMs for additional topologies considered in the experimental searches (T5WW, T5ZZ and T6WW for the ATLAS analysis; T1bbbb, T1tttt, T2bb, and T2tt for the CMS one) 
giving the sets listed in Table~\ref{tab:homegrown}.  

For each signal topology, we simulated 10,000 events per parameter point, with the number of parameter points per topology ranging between 251 (CMS-SUS-16-033, T1bbbb) and 2635 (CMS-SUS-16-033 and ATLAS-SUSY-2016-07, TGQ). The total number of parameter points amounts to 22829.
We used MadGraph5\_aMC@NLO~\cite{Alwall:2014hca} to simulate  the hard scattering (with one additional hard jet) processes and  
Pythia\,8~\cite{Sjostrand:2014zea} for the decays and parton shower, employing the MLM scheme for matching and merging. 
The events were then subjected to the MadAnalysis\,5 
framework, which uses Delphes\,3~\cite{deFavereau:2013fsa} for emulation of the detector response. 
The concrete recast codes used were \cite{ma5recast:altas} and \cite{ma5recast:cms} for the ATLAS and CMS searches, 
respectively, each with its specific Delphes\,3 configuration. 
Note that \cite{ma5recast:cms} employs the aggregate regions of \cite{Sirunyan:2017cwe}, which (as also mentioned in the CMS paper) gives somewhat weaker limits than the full analysis.
In a final step, the efficiencies and their relative errors were read from the MadAnalysis\,5 output and adapted for \mbox{SModelS} to form a total of 351 individual EMs (22 signal regions $\times$ 10 topologies for ATLAS-SUSY-2016-07 and 12 aggregate regions $\times$ 11 topologies for CMS-SUS-16-033, minus one which has only zero efficiencies for one topology in one region).

For the bulk of the EMs, the statistical uncertainty is typically of the order of 1--10\% for the leading topologies in accordance with the accuracy of the recasting with MadAnalysis\,5. However, it may go up to order 30--50\% in some cases, in particular for very low masses or small mass splittings, where also the Monte Carlo-based recasting is less accurate. Home-grown EMs based on higher statistics will be provided in future releases.

\section{Physics impact}
\label{sec:reach}

\begin{figure*}[p]\centering
\includegraphics[width=0.42\textwidth]{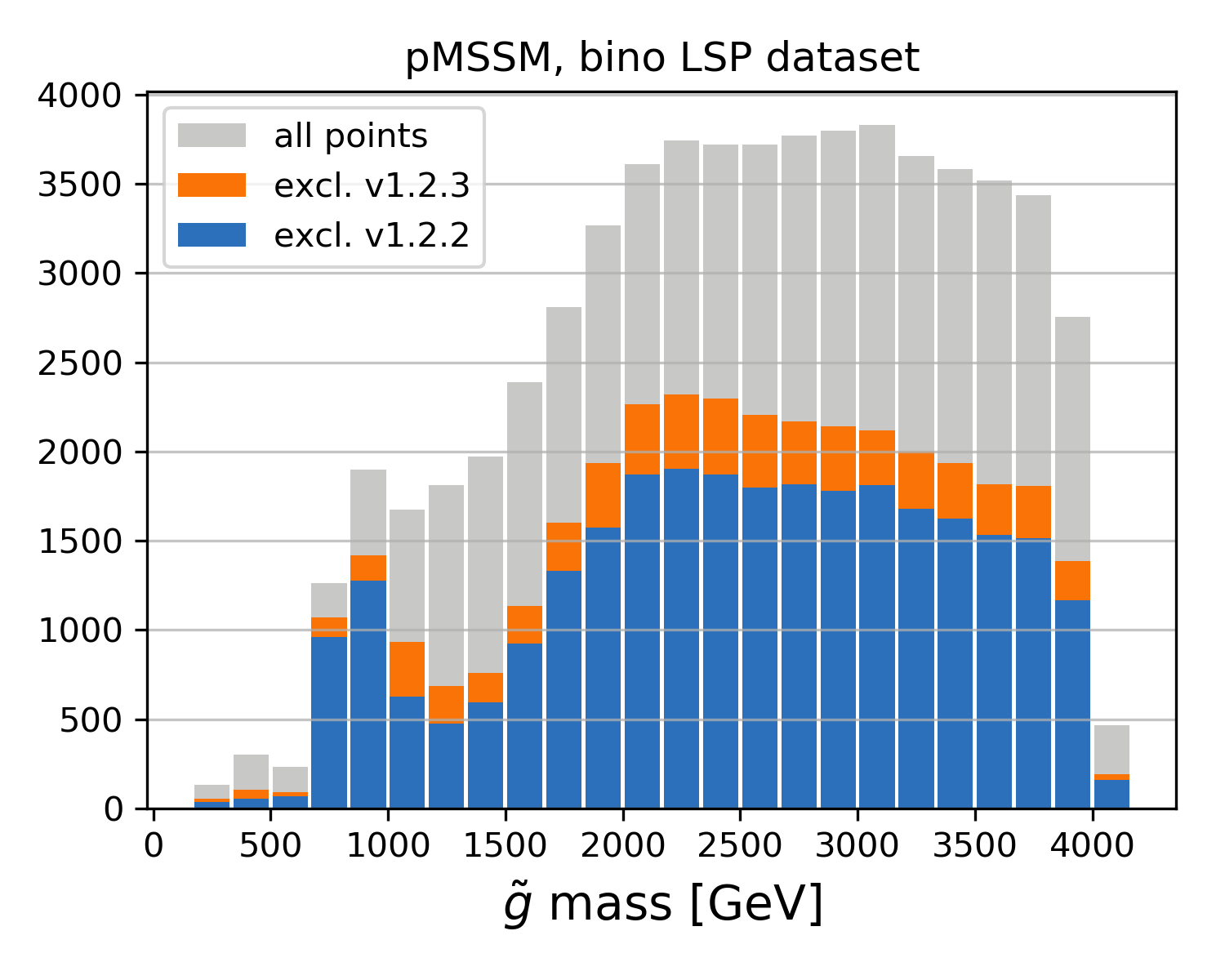}\includegraphics[width=0.42\textwidth]{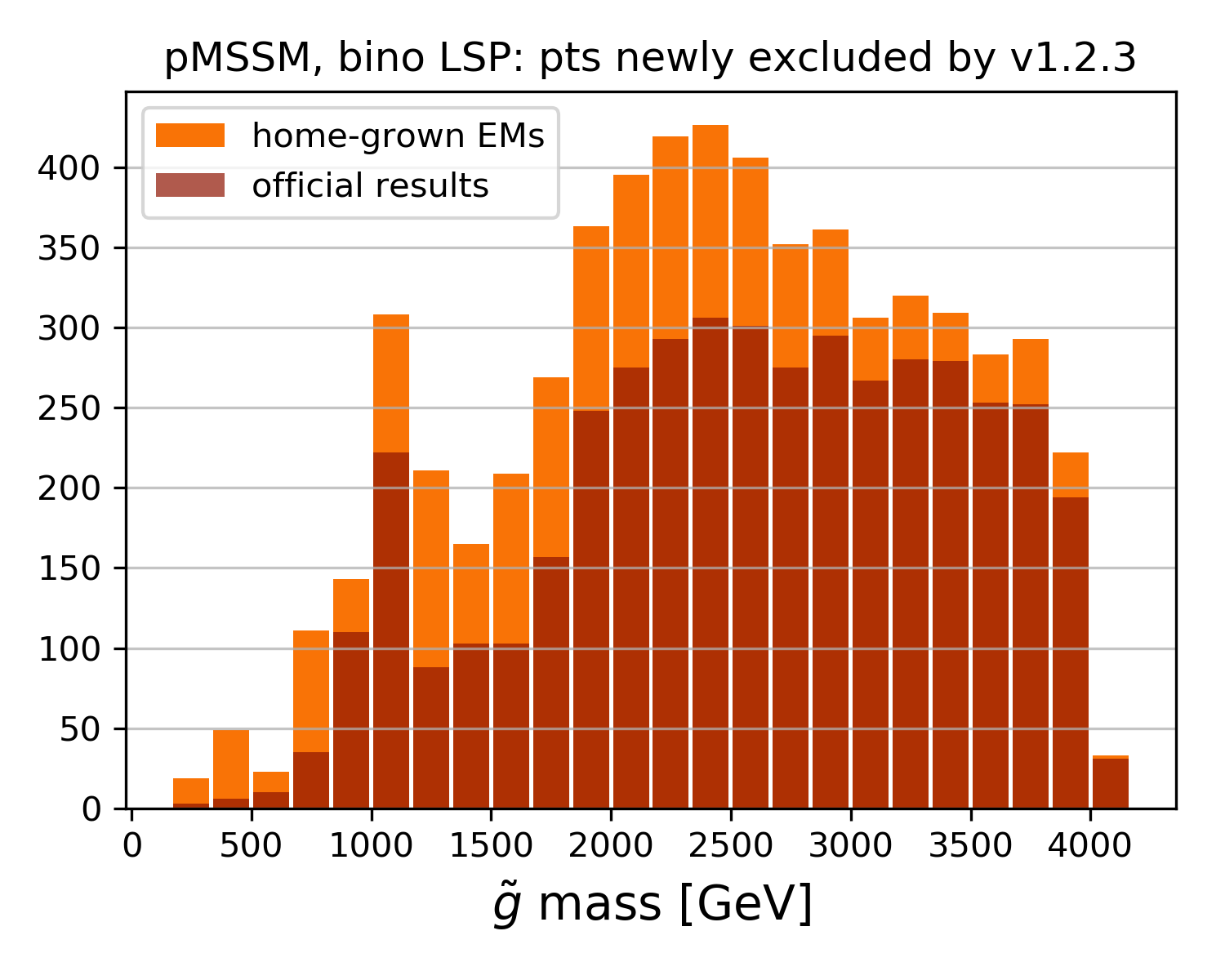}\\
\includegraphics[width=0.42\textwidth]{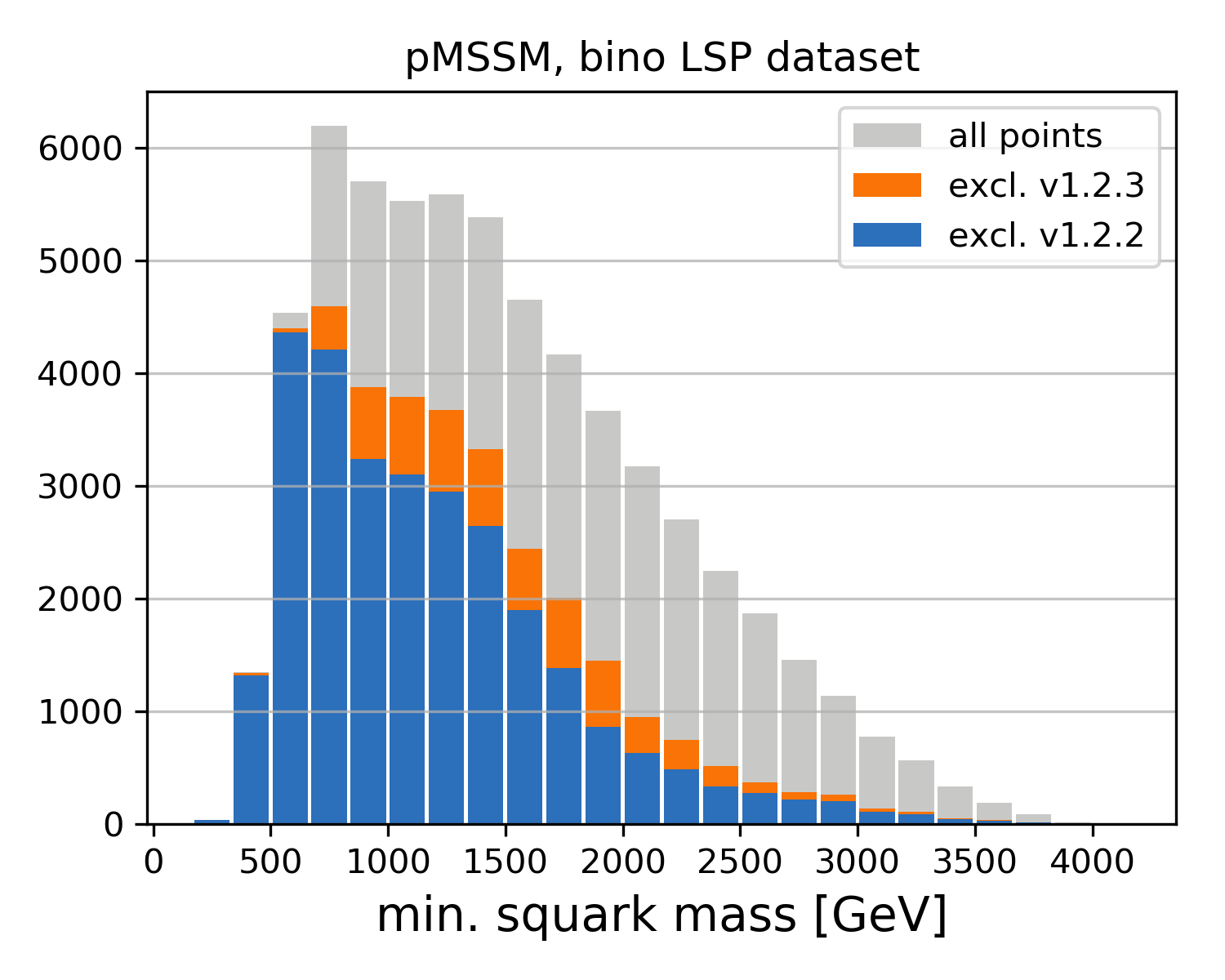}\includegraphics[width=0.42\textwidth]{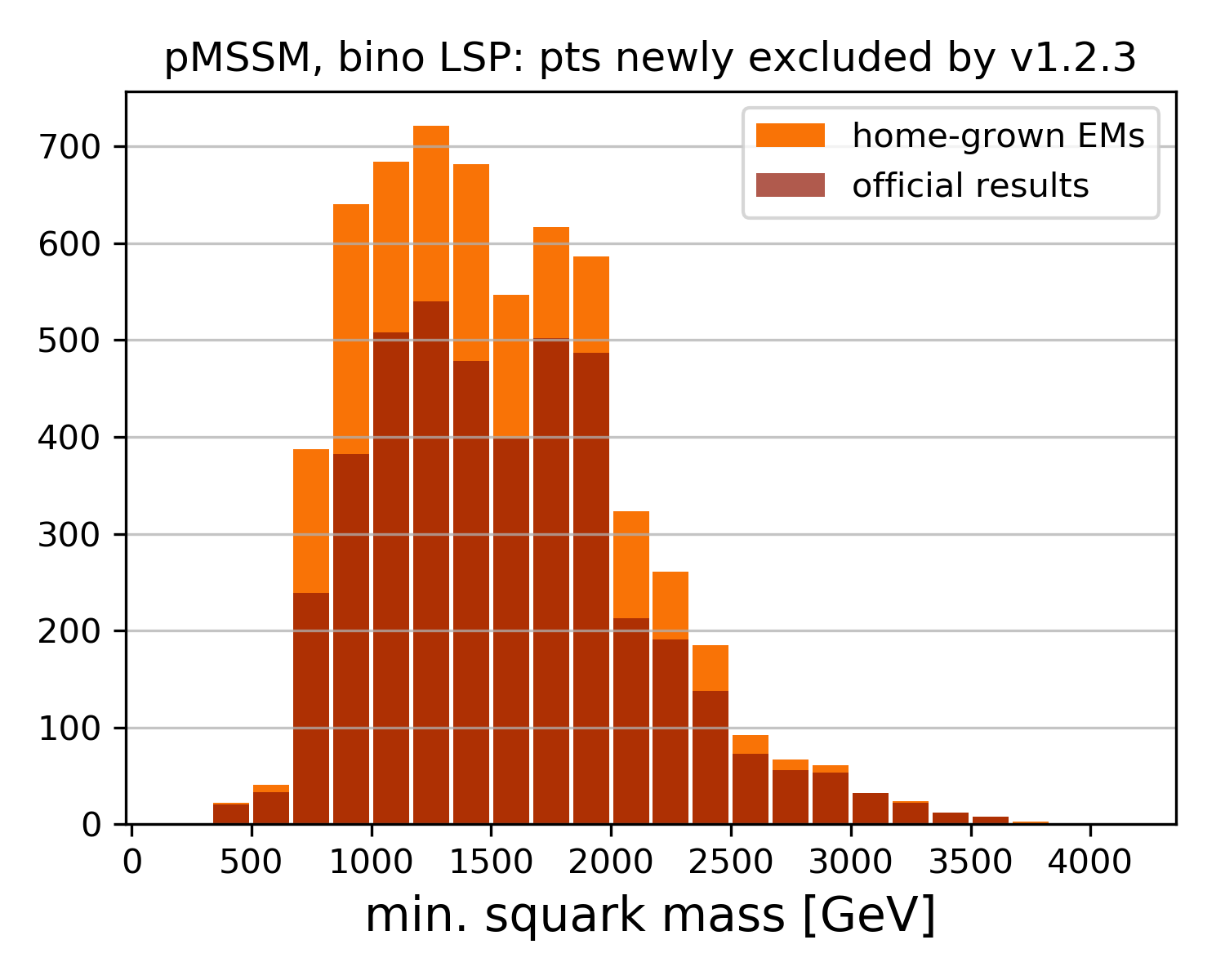}\\
\includegraphics[width=0.42\textwidth]{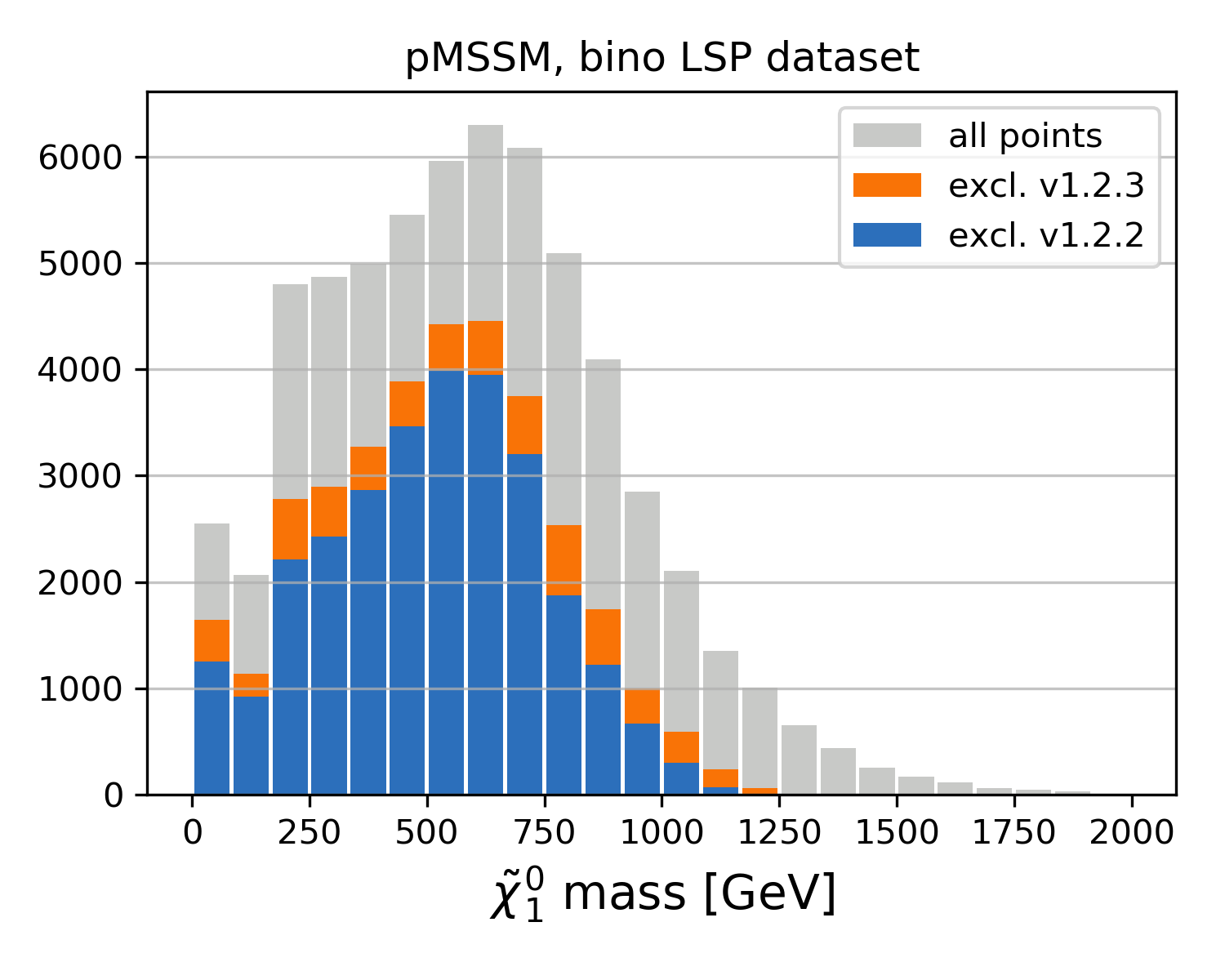}\includegraphics[width=0.42\textwidth]{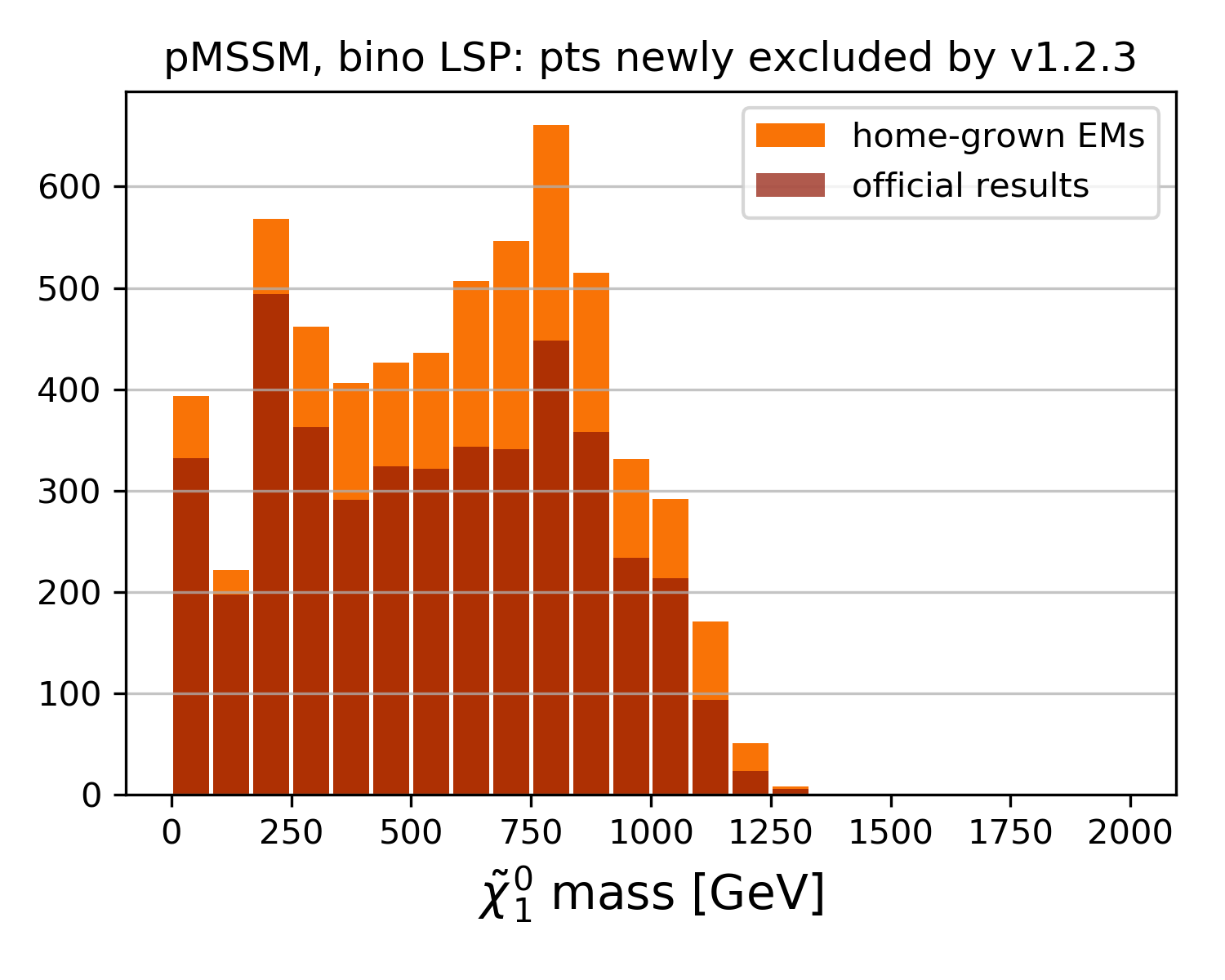}
\\
\caption{\label{fig:pmssm}
Results for the ATLAS pMSSM points with a bino-like LSP, which were classified as not excluded in \cite{Aad:2015baa}, as function of the mass of the gluino (top row), lightest squark (middle row), and $\tilde\chi^0_1$ LSP (bottom row). 
The histograms on the left show the number of points excluded by SModelS v1.2.2 (blue) and the number of additional points excluded with the v1.2.3 update (orange); the distribution of all points in the dataset is shown in grey.
The histograms on the right detail the number of newly excluded points when using only the official ATLAS and CMS results in v1.2.3 (dark red) as compared to using the full v1.2.3 database including the home-grown efficiency maps (orange).}
\end{figure*}

We demonstrate the increase in constraining power of SModelS owing to this database update upon the minimal supersymmetric standard model (MSSM) with 19 free parameters defined at the weak scale---the so-called phenomenological MSSM (pMSSM). To this end, we make use of the extensive dataset from the ATLAS pMSSM study~\cite{Aad:2015baa} available at~\cite{ATLASpMSSMhepdata}. 
Concretely, we use the ATLAS pMSSM scan points with a bino-like neutralino as the lightest supersymmetric particle (LSP), which were not excluded by ATLAS at 8~TeV. 
This amounts to about 61.4K points with sparticle masses up to 4~TeV, of which 28.4K (46\%) are excluded by SModelS~v1.2.2. This increases to 34.4K excluded points (56\%) with v1.2.3.\footnote{For comparison, with the update from v1.1.2~\cite{Dutta:2018ioj} to v1.2.2 the number of excluded points  increased only by about 2\%.} 
Of the 6K newly excluded points, 27\% are excluded only by the home-grown EMs from Table~\ref{tab:homegrown}. 
Figure~\ref{fig:pmssm} shows the number of excluded points as functions of the gluino, lightest squark, and LSP masses. The histograms on the left illustrate the increase in constraining power from SModelS v1.2.2 to v1.2.3, and the histograms on the right show the relevance of the home-grown EMs in v1.2.3. 

SModelS reports its results in the form of $r$-values, defined as the ratio of the theory prediction over the observed upper limit, for each experimental constraint that is matched in the database. All points for which at least one $r$-value equals or exceeds unity ($r_{\rm max} \ge 1$) are considered as excluded. It is instructive to see which analyses/results, thus, turn out as the most constraining ones.  
For the 6K points newly excluded with v1.2.3, the highest $r$-value comes in 51\% of the cases from CMS-SUS-19-006 (mostly the T2 ULs), followed by ATLAS-SUSY-2016-07 EMs (36\%) 
and ATLAS-SUSY-2018-32 (6\%, mostly TSlepSlep ULs). 

The importance of the combination of contributions from the topologies shown in Figure~\ref{fig:EMbakery} is best shown upon an explicit example. 
To this end, we choose {\tt 424686577.slha} from the ATLAS pMSSM dataset~\cite{ATLASpMSSMhepdata}. 
This point features heavy squarks ($m_{\tilde{q}} \sim 2.5$--$3.5$~TeV), a light gluino ($m_{\tilde{g}} \simeq 976$~GeV), and an LSP which is close in mass to the gluino ($m_{\tilde\chi^0_1} \simeq 908$~GeV).  
The gluino decays to the LSP via two competing modes, 3-body decays into the LSP plus two quarks (${\rm BR}\simeq 40\%$ for $q\bar q\tilde\chi^0_1$, 16\% for $b\bar b\tilde\chi^0_1$) or loop decays into the LSP plus a gluon (${\rm BR}\simeq 44\%$). 
As a result, the signal of gluino-pair production consists of  
16\%~T1, 19\%~T2, and 35\%~TGQ.
The heavy squarks, on the other hand, decay into the gluino plus a quark with BRs of 90--99\%. Thus, gluino-squark associated production generates the T3GQ and T5GQ topologies from Figure~\ref{fig:EMbakery}
(the contribution from pair production of squarks is subdominant). 
Table~\ref{tab:SMSsummary} lists the individual contributions of each topology to the total signal for the best signal region of the CMS-SUS-16-033 analysis. As we can see, each individual topology contributes with an $r$-value less than 0.3, and only the combination of all contributions allows SModelS to exclude this particular point. This is only possible due to the efficiency maps listed in Table~\ref{tab:homegrown}. 

\begin{table}
	\tbl{Contributions from each individual SMS topology to the total $r$-value for the pMSSM point {\tt 424686577} discussed in the text. They correspond to the ratio between the signal yield and the observed upper limit for the signal region $N_{jet} = 5, N_b = 0, H_T > 1500~\mbox{GeV}, H_T^{miss} > 750~\mbox{GeV}$ from CMS-SUS-16-033.
		\label{tab:SMSsummary}}{
		\begin{tabular}{ |c|c|c|c| }
			\hline
			{Topology}  &  SUSY process(es)  & 
			\pbox{1.6cm}{ Contribution\\[-1mm] to $r$-value} \\ \hline
			T1 & $pp\to \tilde{g}\tilde{g} \to q q \tilde{\chi}_1^0 + q q \tilde{\chi}_1^0$ & 0.16 \\ \hline
			T2 & $pp\to \tilde{g} \tilde{g} \to g \tilde{\chi}_1^0  + g \tilde{\chi}_1^0$ & 0.18 \\ \hline
			TGQ & $pp\to \tilde{g} \tilde{g} \to g \tilde{\chi}_1^0  + q q \tilde{\chi}_1^0$ & 0.17 \\ \hline			
			T3GQ & $pp\to \tilde{g} \tilde{q} \to g \tilde{\chi}_1^0  + q \tilde{g} \left(\to g \tilde{\chi}_1^0 \right)$ & 0.31 \\ \hline			
			T5GQ & $pp\to \tilde{g} \tilde{q} \to q q \tilde{\chi}_1^0  + q \tilde{g} \left(\to q q \tilde{\chi}_1^0 \right)$ & 0.27 \\ \hline						
			\hline
			Total &$\tilde{g} \tilde{g}$, $\tilde{g} \tilde{q}$ production  & 1.09 \\ \hline
	\end{tabular}}
\end{table}

\section{Usage}\label{sec:conclusions}

The new database presented here is shipped with the SModelS~v1.2.3 package, but it can also be used with the earlier v1.2.x code versions.
The easiest way is by specifying the path to the database URL. 
When using \texttt{runSModelS.py}, this means setting   
\begin{verbatim}
path = http://smodels.github.io/database/official123
\end{verbatim}
in the parameters file. In SModelS~v1.2.3, also the shorthand notation \code{path = official} can be used. 
When writing one's own python main program, one has to set 
\begin{verbatim}
database = Database("...")
\end{verbatim}
where the dots stand for the same database URL as above. 
The official, precompiled pickle file \texttt{official123.pcl}  (680 MB) is then downloaded upon first execution. 
Note that this download is often faster than parsing the text database oneself. 

Users who want to update the text database in an existing SModelS v1.2.x installation, can download the .zip or .tar.gz file from 
\url{https://github.com/SModelS/smodels-database-release/releases}. 
It suffices to put this tarball into the main {\tt smodels} folder and explode it there.
That is, the following steps need to be performed:
\begin{verbatim}
   mv smodels-database-v1.2.3.tar.gz <smodels folder>
   cd <smodels folder>
   tar -xzvf smodels-database-v1.2.3.tar.gz
   rm smodels-database-v1.2.3.tar.gz
\end{verbatim} 
The new database will be unpacked into the {\tt smodels-database} directory, replacing the previous version, 
and the pickle file will then be automatically rebuilt on the next run of SModelS.
For a clean installation, it is recommended to first remove the previous database version. 
If the tarball is unpacked to another location,
one has to correctly set the SModelS database path when running SModelS.
If using {\tt runSModelS.py}, this is done in the {\tt parameters.ini} file.

\section{Conclusions}\label{sec:conclusions}

We presented the update of the SModelS database with the simplified model results from 13 ATLAS and 10 CMS SUSY analyses from Run~2 with 36--139~fb$^{-1}$ of data. This comprises   
76 new official UL and EM results from ATLAS and CMS, supplemented by 21 EM results produced by us using MadAnalysis5. 
These results significantly improve previously available  constraints. 

In total, the SModelS v1.2.3 database now contains 
170 UL and 42 EM results\,\footnote{The 42 EM results consist in fact of 458 individual maps in the different signal regions of 6 ATLAS and 3 CMS analyses.} from 23 ATLAS and 29 CMS analyses at 13~TeV, plus a large number of results for 8~TeV.  
The \mbox{SModelS} package is publicly available and can readily be used to constrain arbitrary BSM models which have a $\mathbb{Z}_2$ symmetry, provided that the SMS assumptions \cite{Kraml:2013mwa,Ambrogi:2017neo} apply. We will continue to update it as new ATLAS and CMS search results become available.

Simplified models have become one of the standard methods for ATLAS and CMS to communicate the results of their searches for new particles. In order to maximize the usefulness of this approach, we encourage the experimental collaborations to provide expected ULs in addition to the observed ones, to provide EMs for all signal regions over the full parameter space of the considered simplified model, and generally to provide results for single SMS topologies rather than a combination of two or more topologies. 
Information enabling the combination of SRs, \i.e., full or simplified likelihoods, is also highly valuable.
Detailed arguments and recommendations for the presentation of simplified model results are elaborated in Section~II.A.5 of the LHC Reinterpretation Forum report~\cite{Abdallah:2020pec}.

\section*{Acknowledgements} 

We thank the ATLAS and CMS SUSY groups for providing a vast number of simplified model results in digital form. 
This work was supported in part by the IN2P3 project ``Th\'eorie -- BSMGA''. C.\,K.\,K.\ acknowledges the support from the Royal Society-SERB Newton International Fellowship (NF171488). The work of A.\,L.\ was supported by the S\~ao Paulo Research Foundation (FAPESP), project	2015/20570-1.
 

\bibliography{references}

\end{document}